\documentclass[manuscript]{acmart}
\usepackage[flushleft]{threeparttable}
\usepackage{multirow}
\usepackage{siunitx}
\usepackage{dcolumn,booktabs}
\usepackage{caption}
\usepackage{subcaption}
\usepackage{xcolor}
\newcolumntype{d}[1]{D{.}{.}{#1}}
\AtBeginDocument{%
  \providecommand\BibTeX{{%
    \normalfont B\kern-0.5em{\scshape i\kern-0.25em b}\kern-0.8em\TeX}}}

\setcopyright{acmcopyright}
\copyrightyear{2018}
\acmYear{2018}
\acmDOI{XXXXXXX.XXXXXXX}

\acmConference[ICTD '22]{International Conference on Information & Communication Technologies and Development}{June 27--29, 2022}{Seattle, WA}
\acmPrice{15.00}
\acmISBN{978-1-4503-XXXX-X/18/06}

\begin{document}

\title{Closed Ranks: The Discursive Value of Military Support for Indian Politicians on Social Media}


\author{Agrima Seth}
\email{agrima@umich.edu}
\authornotemark[1]
\affiliation{%
  \institution{University of Michigan}
  \city{Ann Arbor}
  \state{Michigan}
  \country{USA}
}

\author{Soham De}
\email{soham.de\_ug22@ashoka.edu.in}
\authornote{Both authors contributed equally to this research.}
\affiliation{%
  \institution{Ashoka University}
  \city{Sonepat}
  \state{Haryana}
  \country{India}
}

\author{Arshia Arya}
\email{t-aarya@microsoft.com}
\affiliation{%
  \institution{Microsoft Research}
  \country{India}
}

\author{Steven Wilkinson}
\email{steven.wilkinson@yale.edu}
\affiliation{%
  \institution{Yale University}
  \city{New Haven}
  \state{Connecticut}
  \country{USA}
}

\author{Sushant Singh}
\email{Sushant@cprindia.org}
\affiliation{%
  \institution{Centre for Policy Research}
  \city{New Delhi}
  \state{Delhi}
  \country{India}
}

\author{Joyojeet Pal}
\email{joyojeet@umich.edu}
\affiliation{%
  \institution{University of Michigan}
  \city{Ann Arbor}
  \state{Michigan}
  \country{USA}
}

\renewcommand{\shortauthors}{Seth and De, et al.}

\begin{abstract}
Influencers play a crucial role in shaping public narratives through information creation and diffusion in the Global South. While public figures from various walks of life and their impact on public discourse have been studied, defence veterans as influencers of the political discourse have been largely overlooked. Veterans matter in the public spehere as a normatively important political lobby. They are also interesting because, unlike active-duty military officers, they are not restricted from taking public sides on politics, so their posts may provide a window into the views of those still in the service. In this work, we systematically analyze the engagement on Twitter of self-described defence-related accounts and politician accounts that post on defence-related issues. We find that self-described defence-related accounts disproportionately engage with the current ruling party in India. We find that politicians promote their closeness to the defence services and nationalist credentials through engagements with defence-related influencers. We briefly consider the institutional implications of these patterns and connections.
\end{abstract}



\maketitle

\section{Introduction}
Political science and policy scholars have analyzed the discursive practices used, and the policies implemented \cite{tepe2021populist} by political parties (populist and non-populist) to ensure their re-election. However, not many studies have focused on the Global South. The tactics used by political parties are not universal but are instead community aware (i.e., what draws popular support in one country does not necessarily translate to another country). In this paper, we move the focus of our analysis towards India. Like every other democracy, political parties in India also engage various institutions to amass support. Some of these institutions are typically partisan (e.g., religious bodies, caste associations, trade unions), while others are traditionally non-partisan (e.g., the courts, civil services). One such non-partisan group is the Indian armed forces, where both the veterans and those in active service have largely kept themselves out of political engagements. Likewise, politicians kept themselves out of the defence services' internal organizational decisions and did not draw the force into political controversies. This has arguably been a reason for a distinctive feature of the Indian democracy since 1947 --- avoiding coups and military interference in politics that we have seen in the immediate vicinity of Pakistan and Bangladesh, and well as many other post-colonial states \cite{Wilkinson2015ArmyNation, Mukherjee2020absentdialogue}. 

However, recently political parties have engaged various institutions that may not seem directly linked to politics and governance but yield a considerable social influence like sportspersons \cite{mishra2021sporting} and media celebrities \cite{mishra2021rihanna} into politics. This paper aims to discern how political parties are engaging the armed forces in their politics, allowing for an institutional transformation of the armed forces, hereafter referred to as institutional capture. For this, we analyze the activity of veterans who are active on Twitter. 

For the first time, we map the connections and conversations between Indian political leaders and self-described veterans of the Indian defence forces on Twitter. We find that the current self-described defence-related accounts make posts that strongly favor the current ruling party at the center (BJP). This serves as a broader narrative of the party's nationalist narrative. This is happening at a  time when much effort is being put into building a meta-narrative of the nation around a Hindu ethos, and bringing one of the country's most respected and overtly 'secular' institutions is a critical piece of this endeavor. For our analysis, we make generous use of mention-networks, inferred political leanings and relative frequencies. Through novel visualizations, we examine the role of defence-related influencers in creating networks of polarization on social media around the notion of one party best aligned with the nation's security interests.

\section{Literature Review}

Scholarships on populist parties have focused on how political parties influence the pivotal institutions to gain control of the public narrative. While some institutional changes are overt (e.g., police, which are typically non-unionized and report to political masters), others are more covert and gradual (e.g., educational institutions, courts, which typically have significant independence and tenured employment). Recently scholars have moved towards analyzing how politicians seize control of institutions to ensure that the public narrative is in their favor \cite{chesterley2018populism}. Most of the work analyzing this phenomenon has focused on the relationship between identity politics and institutional capture\cite{sata2020caesarean} and the notion of democratic backsliding more broadly \cite{waldner2018unwelcome}. This work of institutional capture must often be carried out by the induction of certain groups that wield normative power. The work most in-line with our study was done by \cite{tepe2021populist}, who analyzed the transformation in the role of \textit{mukhtars} (the oldest locally elected representative institution in turkey) in the country's politics. Through content analysis of 50 presidential speeches to the \textit{mukhtars}, they studied the themes that President Recep Tayyip Erdoğan used to justify their governance and how they convinced the \textit{mukhtars} to further the party's message to their community.

In our work, we aim to study the institutional capture of India's non-partisan institution of defense services. For this, we focus on the discourse on the social media platform Twitter. Social media has gained significant importance in political communication globally. While some of the western nations were early adopters of social media, the list of nations where technologies like Twitter have become a driving force of elections is practically meaningless to count, ranging in regions and political systems from  Indonesia \cite{budiharto2018prediction}, the Philippines \cite{uyheng2019characterizing}, Malaysia \cite{adanan2018twitter}, and Turkey \cite{gokcce2021syrian}, spanning all of Asia, to Senegal \cite{guennoun2019ahead}, Mauritius, \cite{kasenally2017social}, Nigeria, \cite{zoaka2021twitter}, Kenya,  \cite{ochieng2019examining} in the African continent, to practically every nation-state in the Caribbean and Latin American Regions. At the same time, there have been several political events around various parts of the Global South, in which social media, and often Twitter specifically, have played a precipitating role in political and electoral polarization in various countries, including Indonesia \cite{habibi2019analysis}, Kenya \cite{irimba2020problematising}, Pakistan \cite{masroor2019polarization}.

The relatively unregulated space provided by social media allows politicians to communicate in a way (both content and behavior) that they would not be able to through institutions like broadcast media. This has led to political leaders in various nation-states largely or even exclusively moving their direct communication with the people to social media channels like Twitter  \cite{viscardi2020fake, ruiz2020communications, ahmed2014my, sinpeng2020strong}. In India too, Prime Minister, Narendra Modi had moved to communicating almost exclusively through social media by 2013, cutting professional journalists out of his outreach through his entire electoral campaign, and more or less completely after that as well. The Modi team's investment in social media predated its popularity with the electorate, creating Orkut and YouTube channels, well before Facebook or Twitter had widespread adoption in India \cite{pal2018friendly}. 

The 2014 general elections of India saw a unique use of influencers as a form of moral appeal. Modi's own Twitter account frequently featured casual mentions of conversations with respected people in public life \cite{pal2019making}. Further, the leader's of business community, sportsperson and other celebrities were also engaged to provide legitimacy to his vision\cite{chopra2014big, rodrigues2017social, mitra2021modiwithakshay, rai2019may, chakraborty2018political,mishra2021sporting}. 

Past work on influencers in politics has shown, that they play a role in shaping public narratives \cite{rodrigues2017social}, credibility on the goods and services they use \cite{jin2014following}, as well as building a performative virtual consciousness that acts as a barometer of collective values \cite{velasco2020you}. Celebrities and influencers can be critical in deciding election outcomes \cite{recuero2020hyperpartisanship}, just as well as they have been shown to be effective in amplifying political polarization  \cite{soares2018influencers, dash2021divided}. Studies examining the effect of influencers on political discourse in India have overlooked the broader discursive elements of what it means for a political ideology or entity to get support from a population that enjoys other forms of acceptance in society. \\
The defence services are overwhelmingly seen as a positive force in India \cite{kinsella2001symbols}. Surveys over time by the CSDS \cite{StateofDemocracy2013} and the Center for Strategic \& International Studies (CSIS) have consistently found that Indians have more faith in the credibility of defence forces than any other institution, including of course the traditional media houses \cite{abhishek}. In the period following Modi's election to power in May 2014, there have been shifts in the distance between defence service and civilian authority, most significantly seen in first the engagement of the former Chief of Army Staff General VK Singh into the political establishment as a minister in the cabinet and thereafter in the ascendance of General Bipin Rawat, in an out of turn promotion, to the new position of Chief of Defence Staff \cite{joseph2016questions}. Though, in the recent few years much has been written about the BJP's attempts to appear hawkish, with a strong emphasis on a militarization \cite{srivastava2015modi, kaura2020india} and various journalists have explored the relationship between the defence veterans and politicians \cite{mahalingam,krishnan} no attempt has thus far been made to study systematically study the traces of institutional capture of Indian defence services by politicians.

\section{Data}
This section introduces the setup of our analysis. We discuss aggregation of the defence and politician accounts used in the analysis and the subsequent data collection process.

\subsection{Data Goals}
Our primary goal with gathering the data was to understand the connections between politicians and those who self-identified with being in defence-related jobs. In India, people in active duty do not have self-identified public-facing social media, unless they are doing so in an official leadership capacity. As a result, the majority of the ``individual" accounts belong to people who were formerly occupied in the defence services, or those who were related to people in the defence services - such as parents, spouses, children. Because the posts are in English language, they are likely to be disproportionately reflecting the views of officers and their families.

\subsection{Defence database}
We built this database by iteratively collecting the list of defence accounts. Starting with 15 manually identified seed accounts, we collected 1000 most recently followed accounts for each seed account. Next, we filtered the accounts that did not overlap with the initial seeds. We kept only those accounts whose bio contained defence-related words - `army,' `Army,' `navy,' `Navy,' `Command,' `Defence,' `Commander,' `Lieutenant,' `Cpt,' `Captain,' and `Major' resulting in 266 accounts. These defence-related words were manually selected from the most-frequently occurring words in the description texts of the 1000 most-recently followed accounts for all the 15 seed accounts. Next, for each of these 266 accounts, we collected and filtered the 1000 most recently followed accounts, resulting in a final list of 1600 accounts. These 1600 accounts were then manually inspected, and accounts from countries other than India were removed. The final resulting 1300 accounts were manually classified into four categories - (a) official; these were the accounts belonging to the official organizations of the Indian defence forces, (b) self-asserted veterans; these were the accounts belonging to individuals who claim to have served the Indian defence forces, (c) related to defence; these were the accounts belonging to individuals who claim to be related to those who served the defence forces, i.e., spouse, parents or children and (d) others; these were the accounts belonging to individuals who claim to be analysts, content aggregators, fan pages, etc. These were kept separate from self-asserted veterans and related to defence; since such groups and pages are usually home to very varied and unrelated contents, the views represented cannot be attributed to an identifiable user. Especially for this analysis, these pages have no direct connection with the defense personnel or organization.

\subsection{Politician Database}

We use a previously published Indian politicians database built using an ML-based classification pipeline - NivaDuck, that was developed to identify politicians on Twitter in a given country \cite{panda2020nivaduck}. This dataset was used since it probably is the most comprehensive large-scale database of political actors in India ( \textasciitilde30,000 politicians). Starting with a random sample of manually curated Twitter handles of Indian politicians, this database was built by iteratively identifying politicians based on the profile description text and tweet content.

We collect and analyze tweets for each of the accounts in the defence and politician databases for 1 January ’21 - 25 October ’21. We select this time-interval for our study to include major recent events such as the Farmers' Protests, the second COVID wave and the ongoing vaccination campaign and numerous state-level elections. Our methods can be naturally adopted on any timeline for future work.

\section{Analysis}
    
    
    
    

\subsection{Mentions}

An initial analysis of the dates and content of mentions on Twitter shows that Twitter handles of the officials who work directly with the defence services have the highest mention frequency. Prime Minister Narendra Modi (872), as well as his office (825), is the most mentioned political account by the defence-related handles, followed by those of the Minister and Minister of State of Defence - Rajnath Singh (415) and Ajay Bhatt (323) as well as the Defence Ministry Handle (219), followed by the handles of the Home Minister Amit Shah (280), and that of the President (249). We visualise these accounts, as well as the most called-out defence accounts in Figure \ref{fig:callouts}. We observe that the overwhelming majority of these accounts (coloured orange) are leaning towards the current ruling party (BJP) and only a select few defence-veterans (Lt Gen H S Panag, Naveep Singh etc) lean towards the INC.

\begin{figure}[H]
    \centering
    \includegraphics[height=0.35\textwidth, width=\columnwidth]{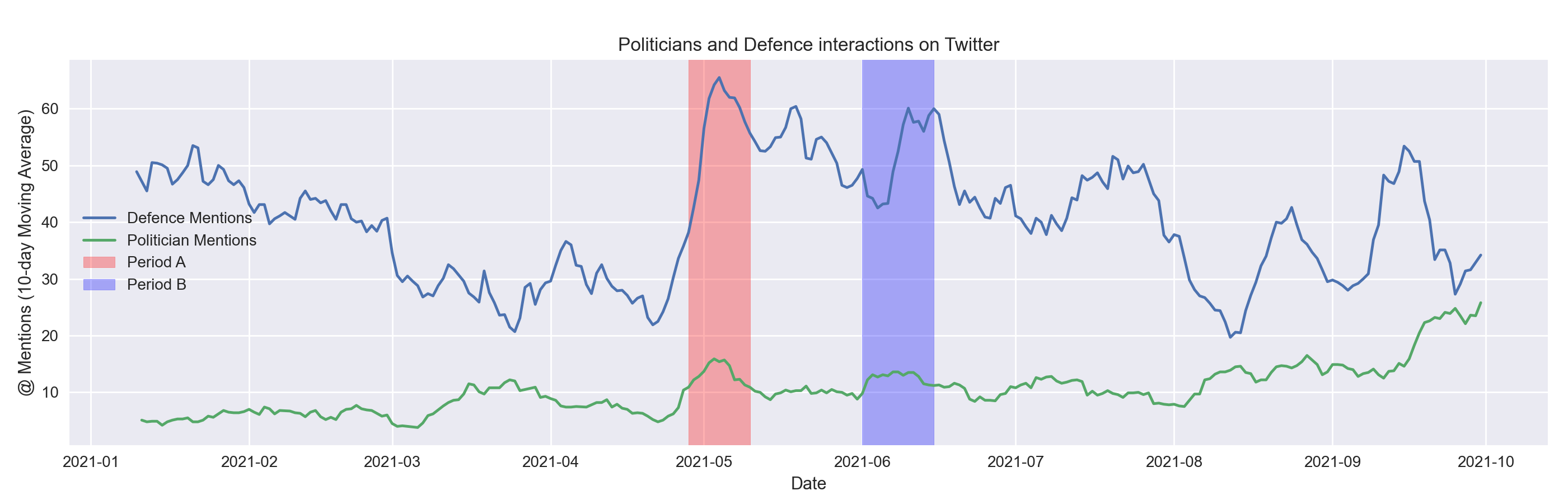}
    \caption{\textbf{Twitter Mentions over time }(10-day rolling average). Blue line represents the count of mentions of defence veteran accounts whereas green represents mentions of politicians. We observe common peaks across both green and blue trend-lines which we discuss further}
    \label{fig:avg}
\end{figure}

\begin{figure}[!htb]
\minipage{0.5\textwidth}
  \includegraphics[width=\linewidth]{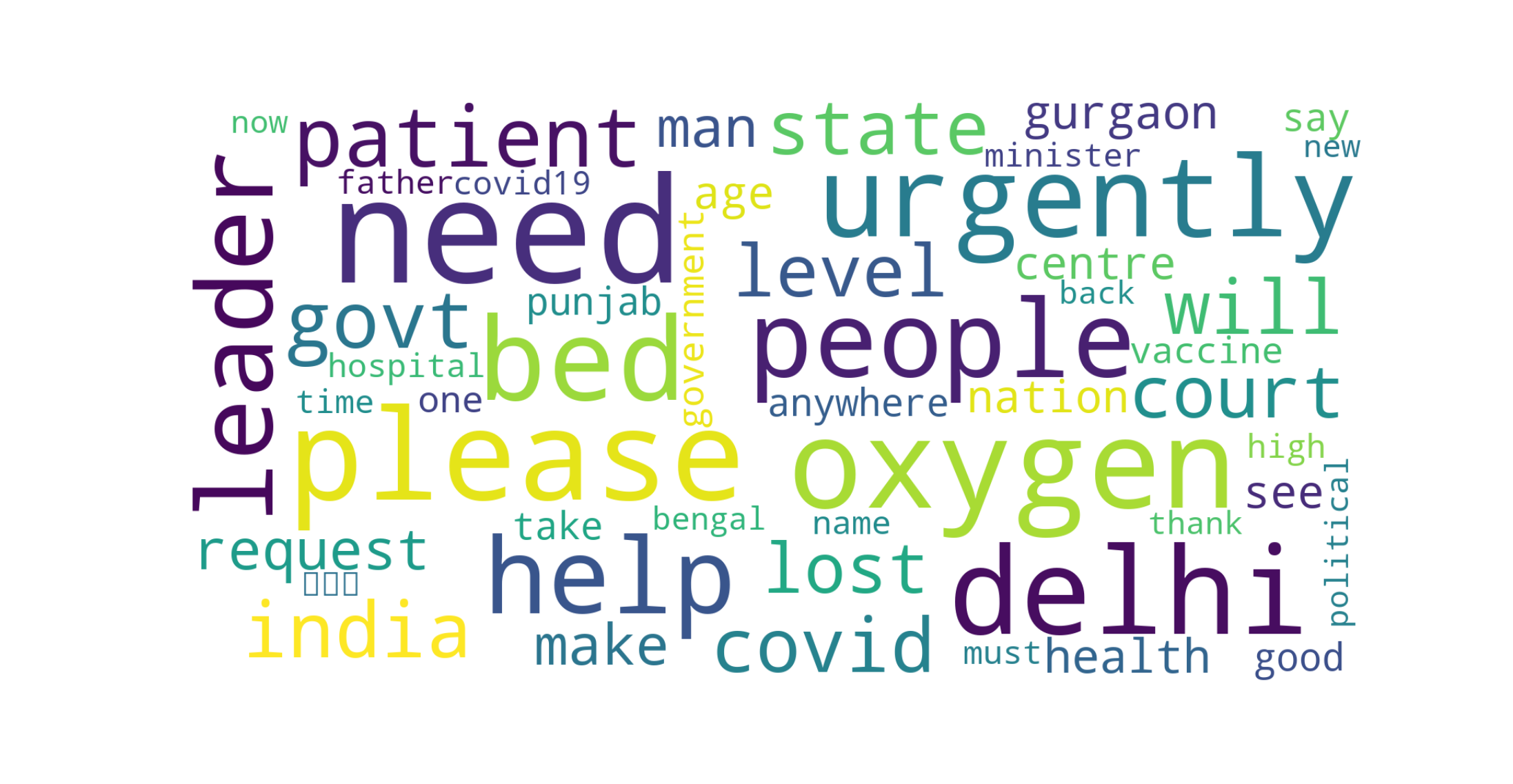}
  \caption{\textbf{Veterans' tweets mentioning politicians.} \\A world-cloud of tweets from veterans during the first peak\\ (April-May)}\label{fig:april_may_politician_mention}
\endminipage\hfill
\minipage{0.5\textwidth}
  \includegraphics[width=\linewidth]{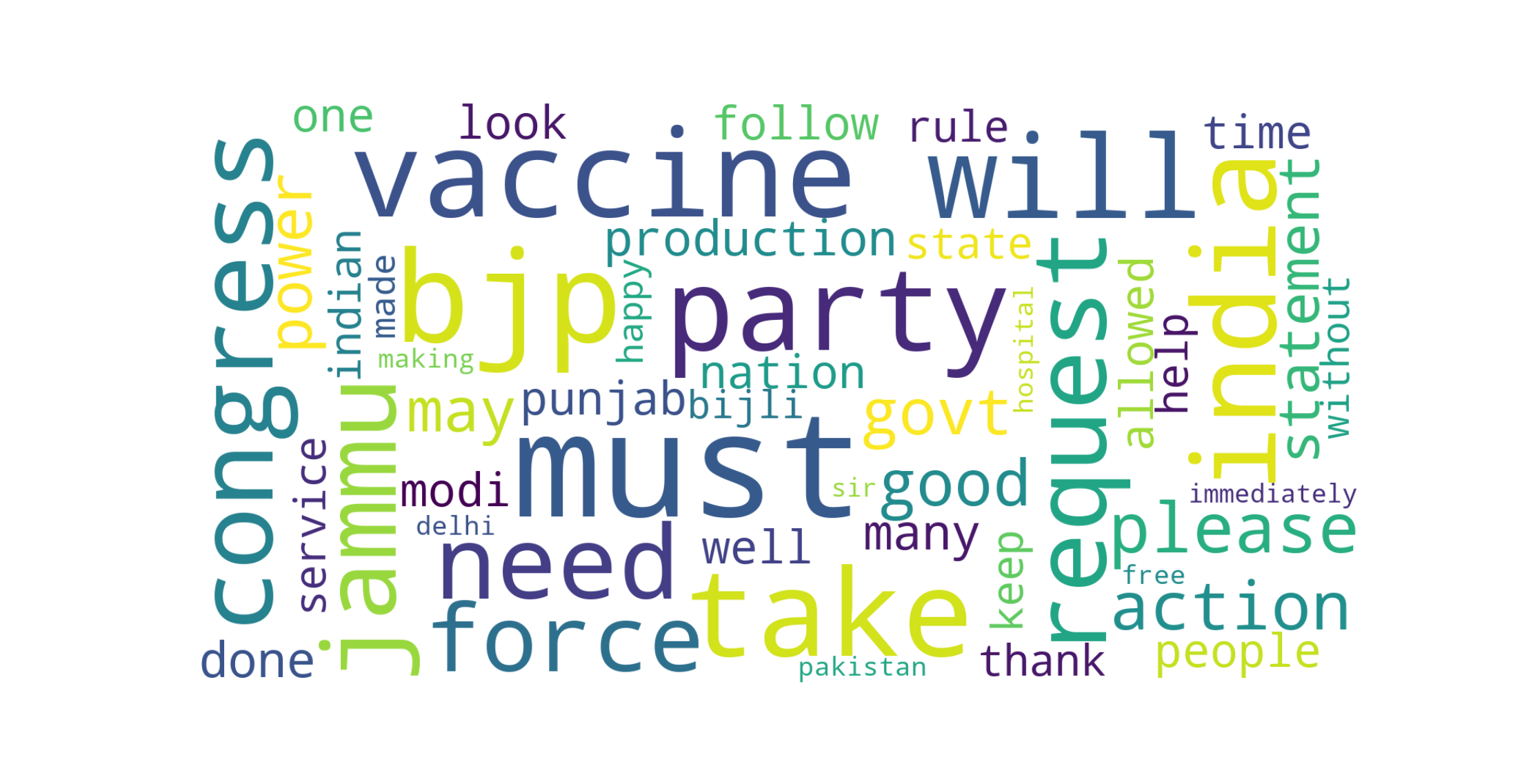}
  \caption{\textbf{Veterans' tweets mentioning politicians.} A world-cloud of tweets from veterans during the second peak (June)}\label{fig:june_politician_mention}
\endminipage\hfill
\end{figure}
\begin{figure}[!htbp]
     \centering
     \includegraphics[height = 0.35\textwidth, width=1.01\textwidth]{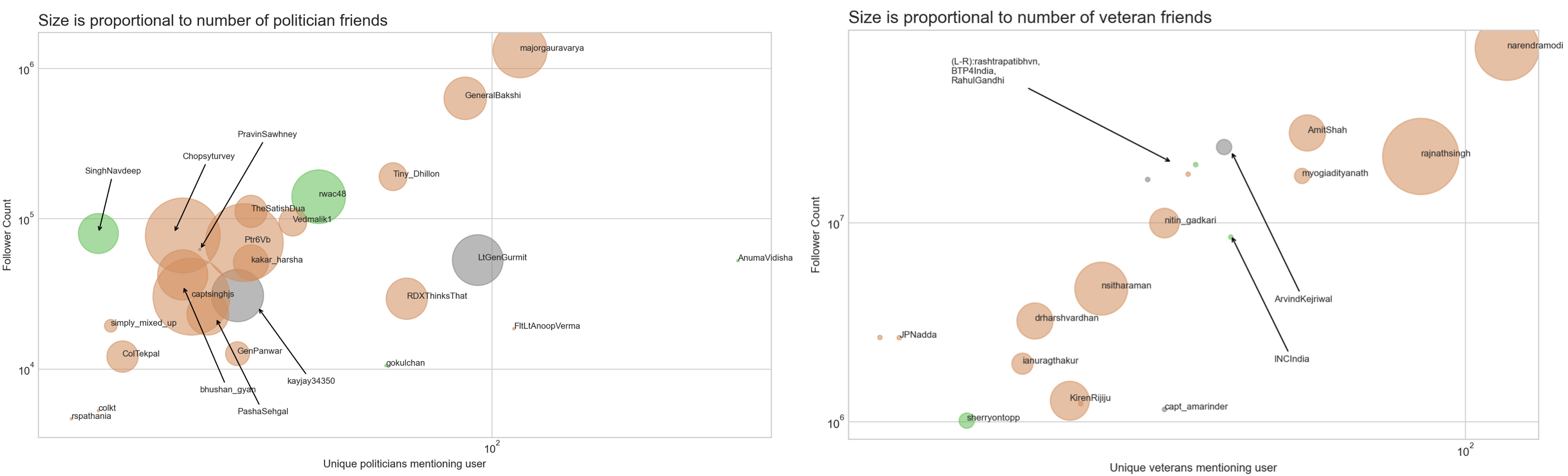}
     \caption{\textbf{Top-25 called-out Veterans (L) and Politicians (R).} Follower Count is represented on the y-axis and the number of mentions is represented on the x-axis. The size of a point is scaled by the total number of tweets made by the account. Orange points lean towards BJP, green points lean towards INC and grey points have ambiguous political leanings}
     \label{fig:callouts}
\end{figure}

We analyze a ten-day rolling average of tweets from politicians' accounts where self-described defence-related accounts were mentioned and self-described defence-related accounts where politicians are mentioned figure \ref{fig:avg}. The first two peaks in 2021 both see a high degree of tweeting from Defence-related accounts aimed at politicians. The first of these is at the peak of the COVID second wave in April 2021. These were mainly panic messages seeking help. This period (Period A, figure \ref{fig:avg}) was marked by an overall increase in Twitter activity as the nation went into a panic over access to medical supplies and hospital beds. As we see in the word cloud of most occurring terms in figure \ref{fig:april_may_politician_mention}, the words need, help, please, and urgently are the most common terms alongside specific asks like oxygen, patient, bed. 

\newpage
During this period, rather than the defence-related accounts reaching out to politicians for defence or defence-related issues, they were using their legitimacy as people who once served in defence to try and get attention to friends and family they were trying to get placed with resources. However, politicians largely avoided this topic during this period when referring to defence-related accounts. Instead, they continued to discuss the farmers' protests. By the second peak (Period B, figure \ref{fig:avg}) from defence-related accounts messaging at politicians in May, where the conversation had shifted to vaccines in figure \ref{fig:june_politician_mention}, as the worst period of the COVID crisis had started subsiding, politicians again chose to talk about Chinese aggressions and the Galwan, avoiding vaccines which were still in short supply.


\subsection{Polarisation}
\begin{figure}[H]
     \centering
     \includegraphics[width=0.85\textwidth]{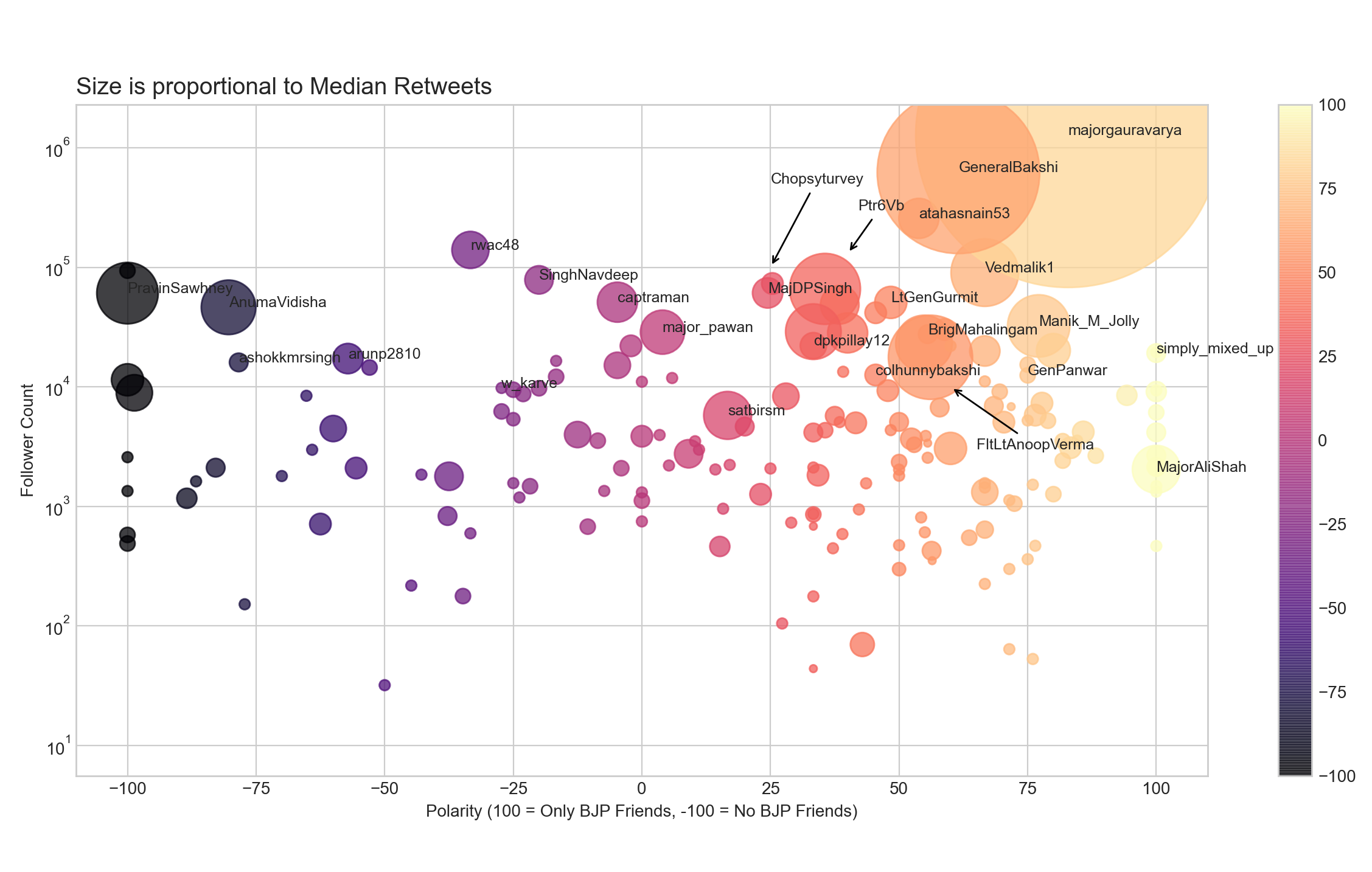}
     \caption{\textbf{Political Polarisation of Defence Users on Twitter.} Polarisation has been calculated based on the party-affiliations of the followed politicians. We observe a higher share of BJP-polarised accounts. Top 20 defence accounts (by in-group following) have been labelled}
     \label{fig:pol}
\end{figure}

We see in figure \ref{fig:pol} that the self-described defence-related accounts are highly polarized towards the BJP, over all other parties combined. This is derived from the fact that the x-axis in figure \ref{fig:pol} is a measure of the fraction of BJP politicians among all the politicians followed by the defence-veteran. The larger bubbles that are higher on both the x and y axes are very highly concentrated on the side of the BJP. To establish the statistical significance of the difference in distribution, we conducted paired t-test. The BJP accounts are followed more significantly compared to non-BJP accounts; $t = 5.7717, p-value = 1.25*10^{-8}$. \\

\begin{figure}[H]
\begin{tabular}{ccc}
  \includegraphics[width=0.32\textwidth]{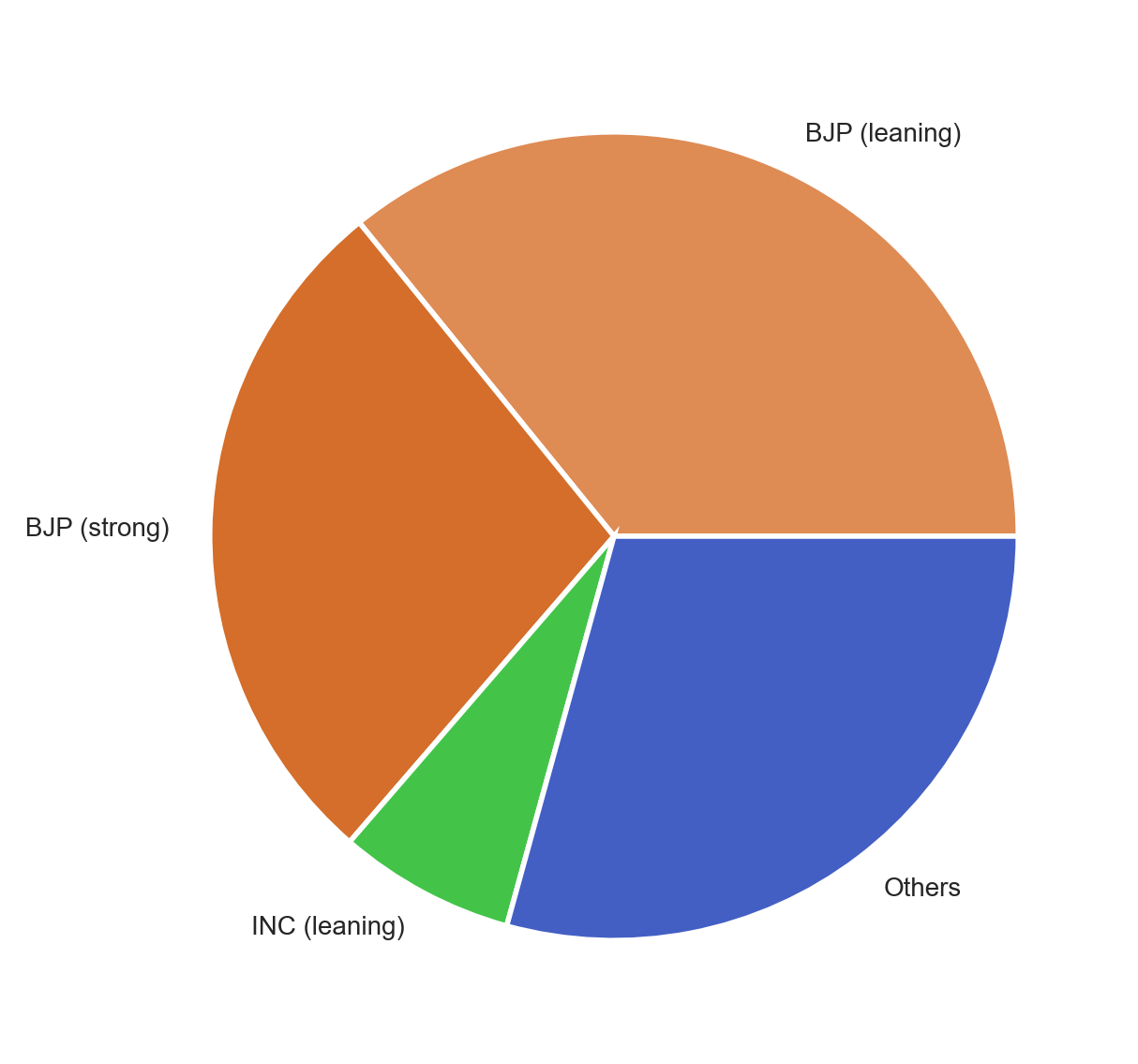} &   \includegraphics[width=0.32\textwidth]{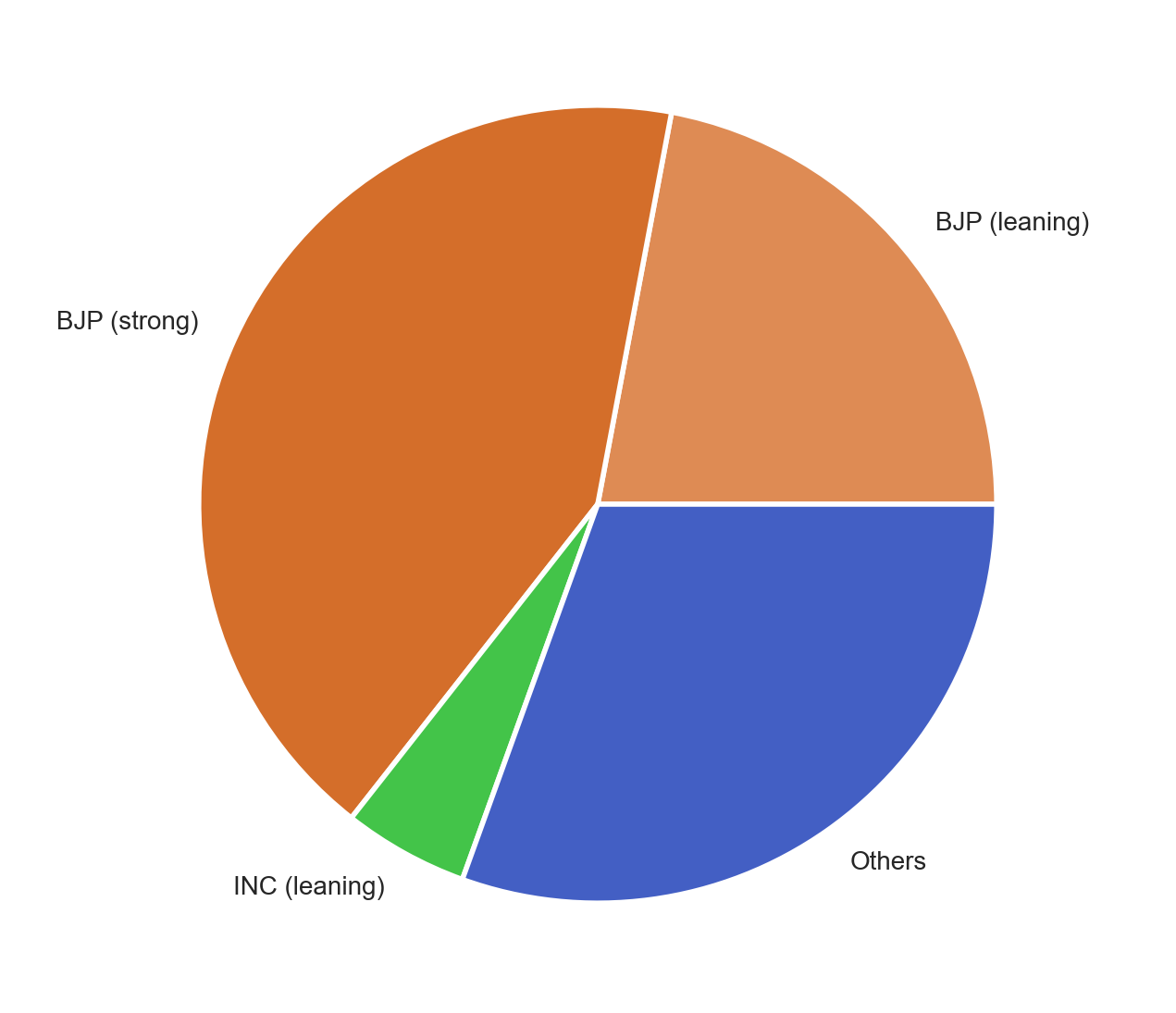} &
  \includegraphics[width=0.32\textwidth]{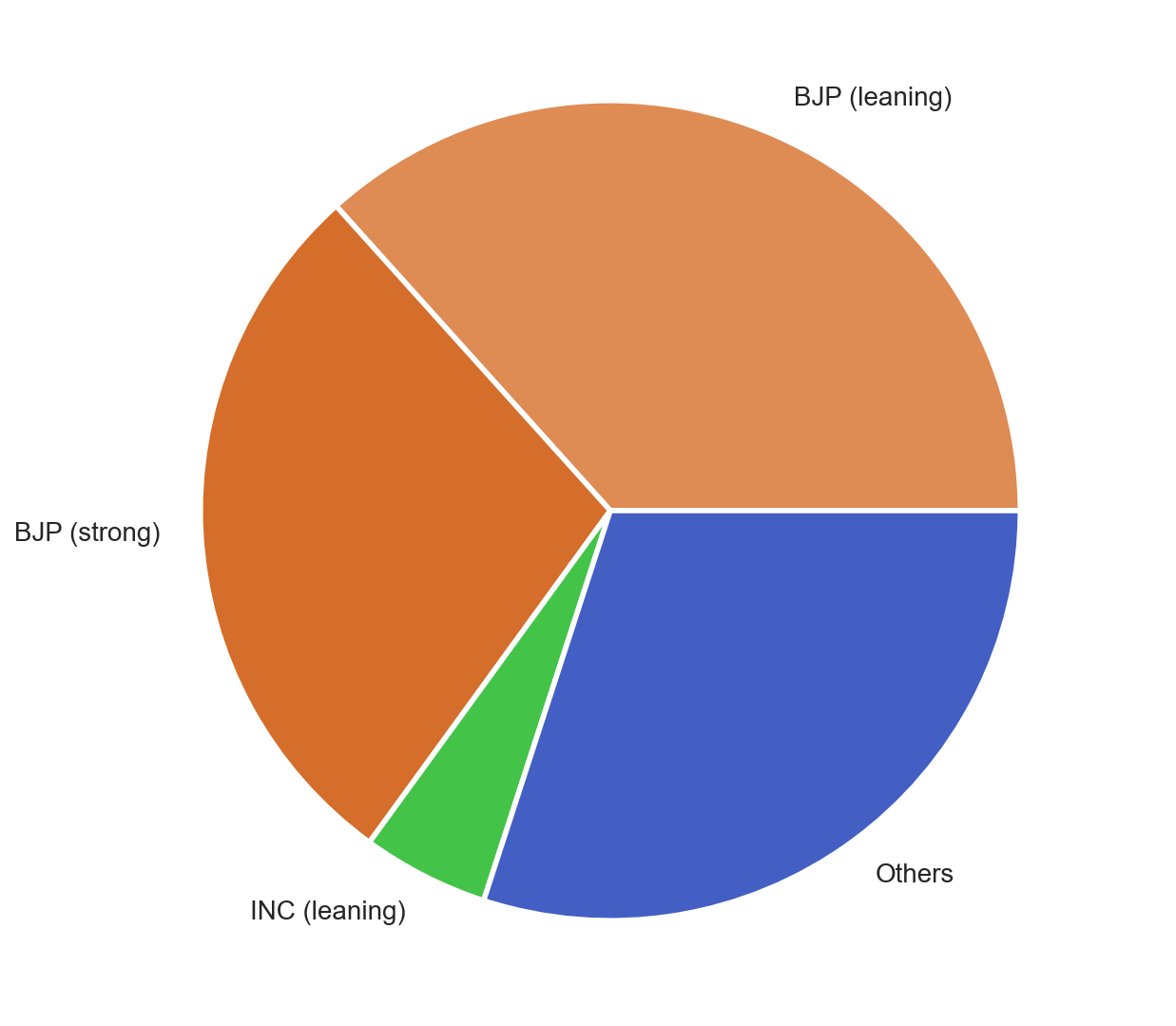}\\
  (a) Self-asserted Veteran & (b) Defence Journalist &
  (c) Related to Veteran \\
\end{tabular}
\caption{\textbf{Political leanings across categories of defence accounts}. Leanings are evaluated based on the party affiliations of the followed politicians}
\label{fig:share}
\end{figure}

Further analysis of self-described defence-related accounts leanings (Figure \ref{fig:share}) shows that majority of these accounts lean towards the ruling government-BJP (as inferred from their following accounts.). Next, we analysed the median retweets of the politician accounts when they tweet about self-described defence-related accounts to their accounts' average retweet rate. The accounts' retweet rate is significantly more when they mention self-described defence-related accounts; $t = 20.492, p-value = 2.2*10^{-16}$. This analysis confirms our primary hypothesis that there is an effective capture of defence-related accounts on the side of the ruling government.

\subsection{Network Analysis}
This section presents the nature and the bi-directionality (or the lack thereof) of the interactions between the self-asserted veterans and politicians on Twitter. To this end, for every veteran, we count the number of times they have tagged a politician on Twitter. Then, for every politician mentioned, we track all the times they have tagged a veteran in their tweets.  \\
\begin{figure}[!htbp]
    \centering
    \includegraphics[height = 0.45\textwidth, width=1.01\columnwidth]{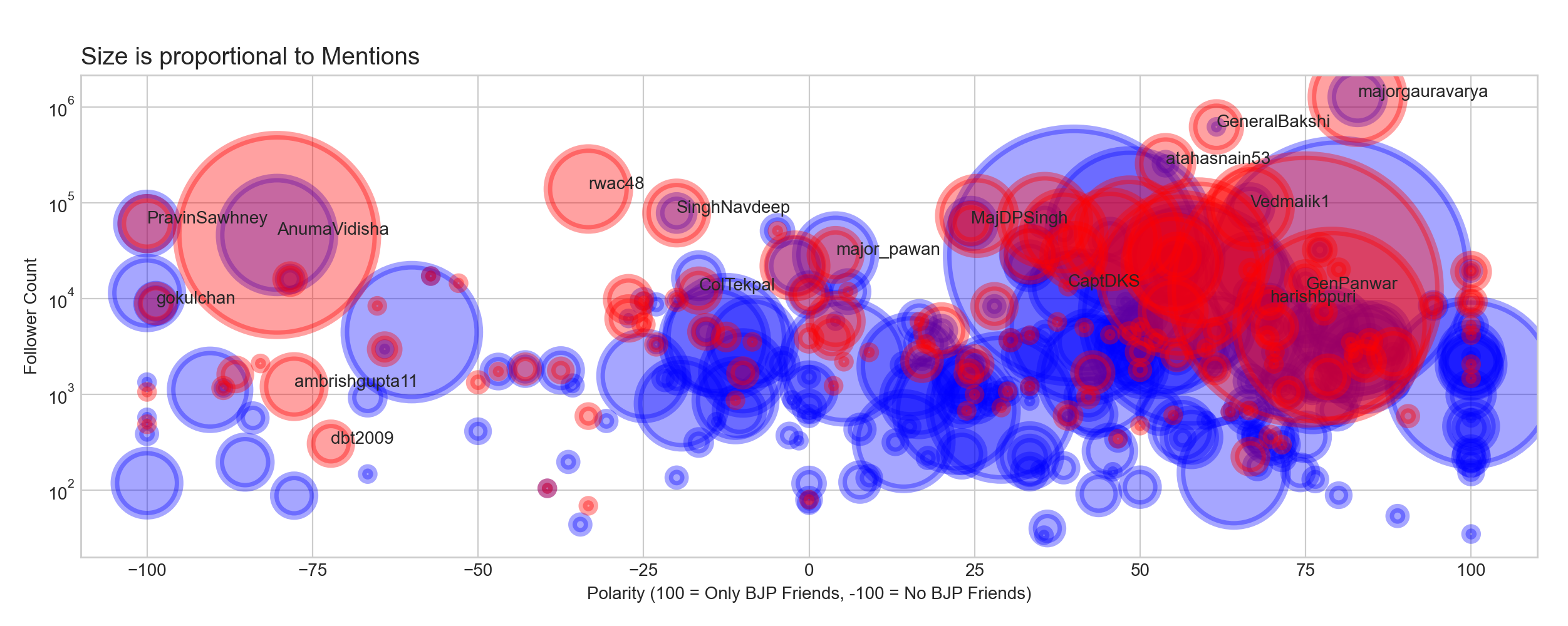}
    \caption{\textbf{Visualising disproportionate bi-directional engagement}. Blue circles represent the frequency of politician mentions by veterans on Twitter and vice-versa. We observe that the interaction between veterans and politicians is largely veteran-led}
    \label{fig:pol_mentions}
\end{figure}


In Figure \ref{fig:pol_mentions}, the blue circles represent the number of times a veteran has tagged a politician. The red circles represent the number of times a politician has tagged a veteran. Therefore, qualitatively, we can define bi-directional engagement as an overlap of red and blue circles. We conducted a linear regression (after removing outliers ,i.e., the data points 3 or more standard deviation away from the mean) to study the effects of polarity and follower count of a self-described defence-related accounts on the engagement that they receive from politician accounts. We make the following remarks (a) Irrespective of the polarity of the account, higher follower count leads to more engagement from politicians' accounts. However, as polarity of accounts increase the effect of follower count on the engagement that they receive decreases $(4.797*10^{-4***})$, and (b) As the follower count of accounts increase the effect of polarity on the engagement that they receive decreases $(-4.039*10^{-6**})$.


\begin{figure}[H]
    \centering
    \includegraphics[width=0.75\columnwidth]{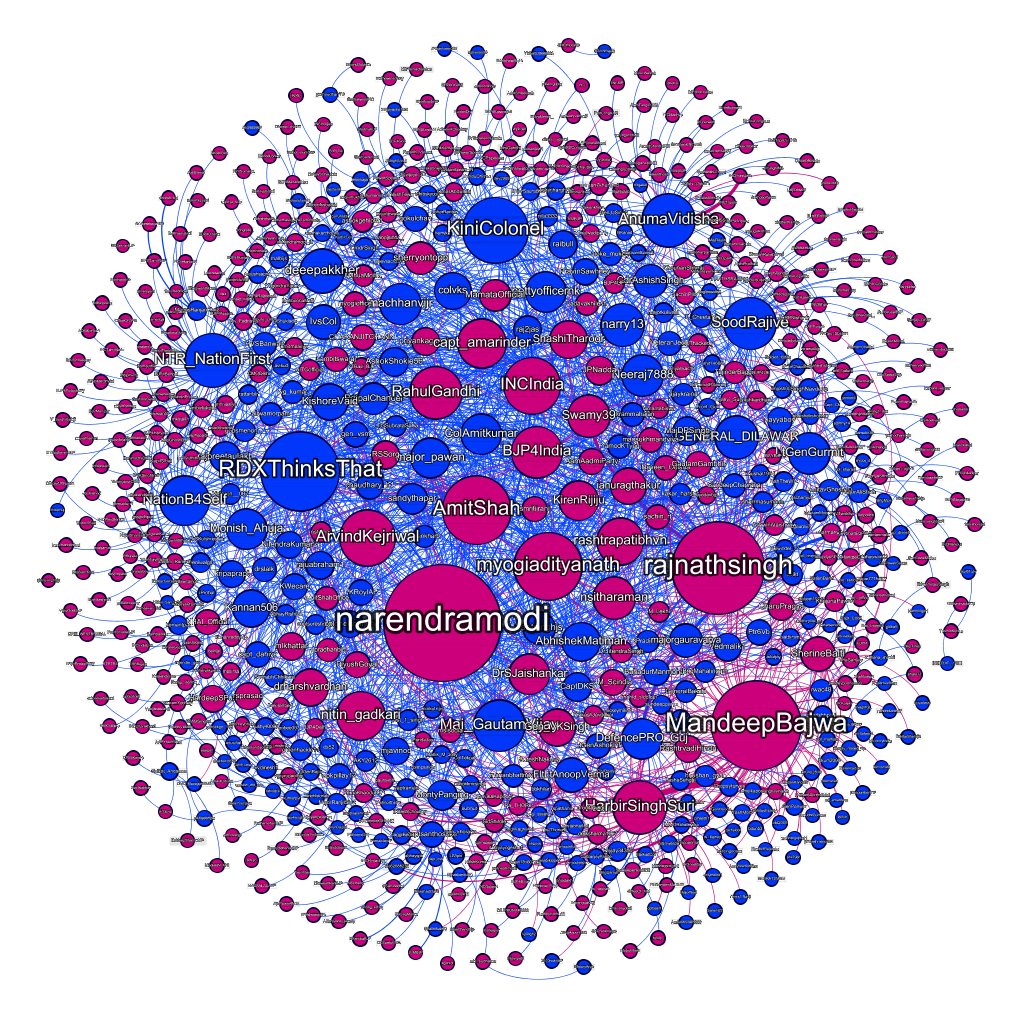}
    \caption{\textbf{Mention networks on Twitter.} Deep pink nodes represent politicians, blue nodes represent self-asserted
    veterans. Deep pink edge $=$ Politician's tweet mentioning defence. Blue edge $=$ Defence persons' tweet mentioning politician. Size is proportional to the node's out-degree.}
    \label{fig:pol_network}
\end{figure}

In Figure \ref{fig:pol_network} we generate the directional connections between self-described defence-related and politicians account. We use the Fruchterman-Reingold \cite{fr} force-directed algorithm in Gephi to generate this visualization. The algorithm treats every edge as a spring and pulls connected nodes together while repelling all others. The deep pink nodes represent politicians whom self-described defence-related accounts have mentioned. The blue nodes represent self-described defence-related accounts that politicians mentioned. A blue edge represents the presence of a tweet from a self-described defence-related account mentioning a politician, and a pink edge represents the presence of a tweet from a politician's account to a self-described defence-related account. The size of a node is scaled proportionally to its out-degree.

Figure \ref{fig:pol_network} shows that the directional connections from defence-related accounts to politicians are spread across politicians, whereas the directional connections from politicians aimed at defence accounts are highly concentrated on a few accounts. Implying that the defence-related accounts also tweet as "citizens," -- meaning they reach out to politicians with problems or send greetings, etc. However, the lines of communication that politicians engage in are highly concentrated in a smaller set of accounts. This suggests that politicians mainly engage with a small number of ``influencers" in the defence space. This shows that the social media cachet of popular defence-related accounts is valuable to politicians. 



\subsection{Qualitative Analysis} 
We selected a small number of highly retweeted messages, or messages from highly followed accounts of key influencers to highlight some of the major themes that emerge in the pool of highly-engaged social media messages. These are meant to offer an illustrative example of the narratives shaping up in the voices of defence-related influencers.

\begin{figure}[H]
\begin{tabular}{cc}
  \includegraphics[width=0.4\textwidth]{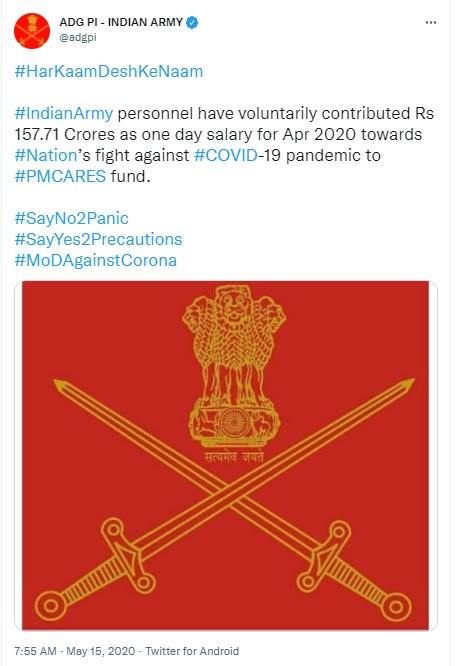} &   \includegraphics[width=0.4\textwidth]{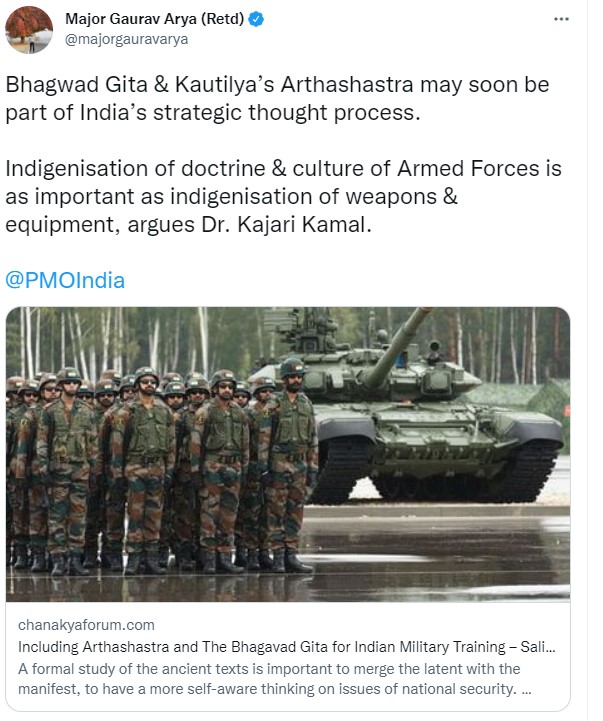}\\
  (a) ADGPI & (b) Maj Gaurav Arya \\

\end{tabular}
\caption{\textbf{Select screenshots from Twitter}}
\end{figure}

The first example, a tweet from the official handle of the Additional Directorate General of Public Information (ADGPI), proclaims that members of the armed forces donated a day's salary to the PM Cares fund. Two things about this tweet are interesting. First, as it would later emerge, the members of the armed forces were not consulted in the decision to make the donation. Second, and more importantly, the fund donated to was not the official government audited fund - the Prime Minister's Disaster Relief Fund, but a separate fund with its own audit chain that has been at the center of much controversy on account of the opacity of its use, and the centrality of the Prime Minister's persona around the fund. The fund was popularized through a series of social media messages by celebrities. The ADGPI's message brought legitimacy to it by underlining that the nation's most respected professionals were willing to make a personal sacrifice for it. 
In Figure 6b, we see a message from Major Gaurav Arya, a defence-veteran and the single most engaged and retweeted defence-related influencer on social media. The reference in this tweet to the engagement of the ancient Sanskrit text on statecraft, Arthashastra, and the Hindu religious text Bhagawad Gita as proposed doctrinal documents for the armed forces is a departure from the traditionally accepted secular ethic in the defence services. While Arya does not represent the government, he is among the most important pro-BJP social media influencers, who also runs his own online magazine - Chanakya Forum. Given that Major Arya is not officially part of the government, his message presents a form of public buoyancy for the idea of a defence culture based on Hindu ideals, and in this, presents a public preference for the notion, which has a different value than if it came directly from the party.

\begin{figure}[H]
\begin{tabular}{cc}
  \includegraphics[width=0.4\textwidth]{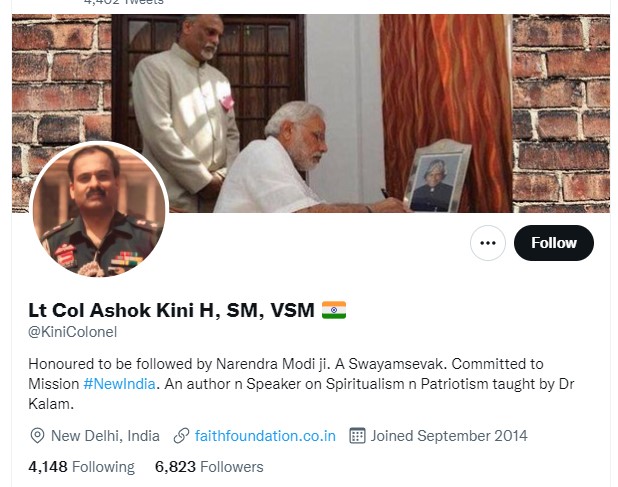} &   \includegraphics[width=0.4\textwidth]{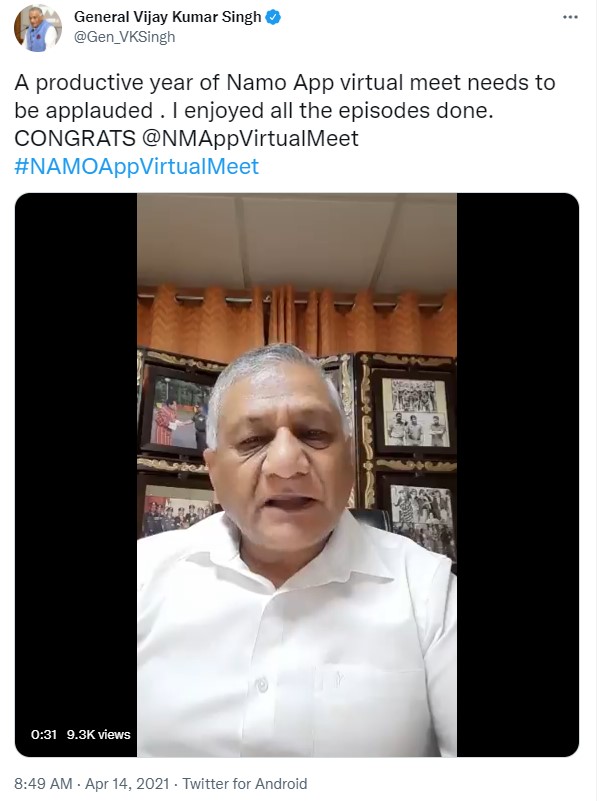} \\
  (c) Lt Col Ashoka Kini & (d) Gen VK Singh \\

\end{tabular}
\caption{\textbf{Select screenshots from Twitter} (contd)}
\end{figure}

In Figure 6c, we see the profile of Ashok Kini, another defence-veteran who takes a strong pro-BJP position in his messaging. In his profile texts, he notes as part of his identity that he is followed by Prime Minister Modi's Twitter handle, something that party workers did right after Modi followed them on Twitter. Indeed, as past work has shown, followbacks by Modi were both provided as a ``reward" for doing the party's work but also served to galvanize the recipient to tweet more aggressively on the party's behalf \cite{pal2016twitter}.

Figure 6d is from General VK Singh, the once Chief of Army Staff, who moved to the BJP after retiring from service. Singh would eventually join the Modi cabinet and serve on several key ministries and has had one of the party's most influential accounts on Twitter. The message here is an endorsement of the NaMo app, named after Narendra Modi, on which VK Singh appears to provide virtual meets to citizens. The endorsement itself is aimed at highlighting not the government's performance as the Minister for Civil Aviation but the act of conducting meetings on the app. Moreover, it is important to note that the NaMo app is an independent app, branded after the Prime Minister and not the party. It is equally interesting that a minister for the government conducts virtual briefings not at a broad social media platform meant for all citizens but on an app that on name and composition is set up for fans of the leader.

\section{Discussion}
The fundamental "development" question we ask is what it means for a political party in power to dominate a subset of the population on social media. We find strong evidence that self-described defence-veteran accounts tend to lean towards the ruling party. To understand this, we should turn to the composition of the armed forces, to understand how the demographics of the defence services may have intersections with the traditional vote-bank of the BJP.

If looked at purely from the prism of caste, religion, and regional origin, the Indian Army JCOs are historically drawn from the norther and western states, as well as the from among upper caste Hindus groups \cite{Wilkinson2015ArmyNation}. These groups have a stronger vote-bank representation within the BJP \cite{mcdonnell2019right}. While the defence services does not hire or discriminate based on caste, the highest echelons of the defence services tend to be over-representation by upper-caste Hindu and Sikh officers -- all of the post-independence Chiefs of Army staff, for instance, are from this demographic. 





So the fact there is a correlation between English-speaking veteran officers and support for the BJP is not that surprising in itself, and going back to the very foundation of the BJP in 1980 we can see individual cases of senior retired officers making statements in support of the party and its aims.
What seems to be distinctive now is that: a) veterans and people from defence families seem increasingly willing to wear their political preferences on their sleeves in social media posts on all kinds of contemporary and sometimes divisive political issues; b) the large number of posts on social media create the perception, unlike the few isolated statements from veteran officers in the past, that a large proportion of the defence services \textit{as a whole} shares the BJP's political views; c) ruling-party politicians for their part are frequently posting on defence issues, and identifying the defence and its actions with their own political agenda.


First, while the defence services' role as a paramount institution that demands respect across democratically elected societies that do not have the draft is relatively universal, a specific post-colonial history matters in India's case. Unlike the colonial era police, who were largely seen as enforcers of the colonial masters' repression, the defence services were outward-facing and carried the reputation of professionalism, modernism, and secular spirit. Also, unlike the police, projected as corrupt and oppressive to the common citizen in the popular imagination, the defence services, outside of the areas where it is used against the nation's own citizens, has largely been perceived as incorruptible self-sacrificing, and reliable. 


Looking at this from a developmental lens, what technology allows is not necessarily an immediate capture of the institution of the defence services but a projection of its perfect alignment with a political vision. This would arguably never be possible without social media since it was not typical, with the exception of a small subset of high-ranking officials or people from already political families who joined the defence services to serve as political endorsements. Twitter has allowed an amorphous group of self-declared defence-related persons, led by a small number of highly-networked influencers, to speak for their professional class.

In practical terms, the relationship between politics and defence is driven by high network influencers. The highest engagements between politicians and defence-related accounts are centered on a few highly influential accounts that are vocal on social media. We see from the data that any tweets that are defence-related in any way from politicians get much more attention on social media than the average messages they tweet about. This means that even outside of the normative legitimacy that the appearance of overwhelming support from military-leaning provides, there is a practical value in the Twitter attention economy of talking about defence.  

The deepest significance of the association of politicians with defence is the experience of the 2019 general elections, in which not only did the ruling government swing the tempo of online attention its way by changing upping the ante on nationalistic appeal and talk about defence through strikes on Pakistan just prior to polling, it changed its entire Twitter campaign, stopping conversations about corruption, and instead moved on to defence. Modi's own campaign added the suffix "Chowkidar," a military-derived term meaning "holder of a post," and asked followers likewise to change their profile names. Tens of thousands followed Chowkidar Narendra Modi to rename themselves on Twitter. 

The politics of representation played out in an account proclaiming being "followed by Narendra Modi" as a badge of honor on Twitter has a different meaning when appearing from a decorated soldier because they are beyond reproach in the public discourse. Likewise, an entire arm of the defence services donating a day's salary for a fund run by the Prime Minister suggests that it is morally appropriate for the average citizen to do. A retired major suggesting bringing religious texts into military training is likewise much more meaningful than a political actor doing the same. Likewise, a former Chief of Army Staff becoming a commentator on an app named after the prime minister underlines the scope of who is on one's team. 

As several studies in the last decade have shown, there is a lot of careful thought put into the social media messaging of the BJP; very little is purely incidental. The party's near-total dominance of defence-related influencers helps underline the notion that it is the best situated to deal with threats to the nation. This not only means that the party then becomes the party of choice whenever the nation is seen as threatened, which in India's case is often in perpetuity as a result of land disputes with both China and Pakistan, but by extension, whenever internal security is projected as threatened, the stamp of approval from the defence becomes a critical card.

In the last decade, the use of social media has gone from something that was used by a small number of elites to being the major driver of electoral outreach in some of the biggest campaigns throughout the Global South. More importantly, social media has created a state of "constant campaign" in which parties perpetually build, refine, and enact their agenda on social media. Speaking about defence has its value, as evidenced in retweets. However, the irony we see here is that politicians do not substantively engage with someone because of their service in the defence services. The best evidence of this is in the April-May 2021 peak in tweets. When defence-related accounts started tweeting for help, the responses either helped a selective few highly followed individuals (which had its own performative impact) or offered no substantive response. Moreover, throughout the study period, the engagement between party politicians and defence-related accounts was largely unidirectional, as we see in figure \ref{fig:pol_network}, essentially only privileging a small number of highly networked veterans. In effect, being 'defence-related' does not enable any kind of special privileges in terms of access to politicians.

\section{Limitations}
The results presented are limited by the scope of the accounts collected in the database. The list of veterans was collected using a mix of manual annotation and keyword search. Hence, it is likely that we may have missed out on some accounts. The overwhelming majority of accounts we found were from officers in the defence services, who are a minority in the overall forces, the overwhelming majority of who are on technologies like WhatsApp. While this skews who speaks for the defence services, it does not change the substance of our findings, since the representative effect of the defence being aligned with the ruling party stands, irrespective of the rank of those speaking for the defence services.

Further, the classification of defence-related accounts was based on the account's username and bio; hence, there might be some false negatives for the accounts that did not explicitly mention affiliation with the defence forces. However, the likelihood of these accounts being targeted by politicians is low, and on addition of these accounts to the database, our results should still hold. Our results are limited to the defence-related accounts on Twitter; there may be other platforms that are used more by defence-veterans.

Finally, 2021 was the second year of the COVID pandemic, and much of the general Tweeting activity and discourse was centered the same. While this does not alter our inferences, we believe that analyzing historical tweets from a pre-pandemic era might further reveal interesting trends which may have been masked by the overwhelm of COVID-related panic tweets this year. 


\section{Conclusion and Future Work}

The case of defence and politics intersecting in India highlights a deeper story about institutional capture and how it plays out in the public eye. Institutional capture typically involved an iron fist. Social media has enabled a persuasive, consistent flow of engagement in which the stakeholders become the message. 

Our work is a call to two fields of work that can be very helpful in understanding this moment in the evolution of political communication. On the one hand, we need to think of new ways of understanding the networks and flows of power in the enactment of power in the public discourse. Social media has been studied for its instrumental purposes, and largely in the Global North, this a moment that underlines the representative power of Twitter. On the other hand, this is a call to extend studies of military-state relations beyond its emerging technology-enabled avatars through social media. While the tensions between civil and defence-services will resolve themselves through complex Realpolitik equations, the performance of military-state harmony has just found new frontiers for academia to catch up with.

\section{Appendices}
The data is uploaded and available at \url{https://github.com/actuallysoham/ICTD-Defence-Public}

\begin{acks}
We thank the reviewers for their valuable comments that helped sharpen the focus of this work.
\end{acks}

\bibliographystyle{ACM-Reference-Format}
\bibliography{sample-base}

\appendix

\end{document}